\documentclass{tlp}
\submitted{}
\revised{}
\accepted{}

\input epsf
\usepackage{color}
\usepackage{url}

\newcommand{\ignore}[1]{}

\begin{document}

\long\def\comment#1{}

\title{Efficient Tabling of Structured Data with Enhanced Hash-Consing}

\author[N.F. Zhou and C. T. Have]
{Neng-Fa Zhou \\
CUNY Brooklyn College \& Graduate Center \\
zhou@sci.brooklyn.cuny.edu 
\and Christian Theil Have \\
Roskilde University \\
cth@ruc.dk
}

\pagerange{\pageref{firstpage}--\pageref{lastpage}}
\volume{\textbf{10} (3):}
\jdate{March 2002}
\setcounter{page}{1}
\pubyear{2002}

\maketitle

\label{firstpage}
\begin{abstract}
Current tabling systems suffer from an increase in space complexity, time complexity or both when dealing with sequences due to the use of data structures for tabled subgoals and answers and the need to copy terms into and from the table area. This symptom can be seen in not only B-Prolog, which uses hash tables, but also systems that use tries such as XSB and YAP. In this paper, we apply hash-consing to tabling structured data in B-Prolog. While hash-consing can reduce the space consumption when sharing is effective, it does not change the time complexity. We enhance hash-consing with two techniques, called {\it input sharing} and {\it hash code memoization}, for reducing the time complexity by avoiding computing hash codes for certain terms. The improved system is able to eliminate the extra linear factor in the old system for processing sequences, thus significantly enhancing the scalability of applications such as language parsing and bio-sequence analysis applications. We confirm this improvement with experimental results.
\end{abstract}


\section{Introduction}
\label{sec:Introduction}
Tabling, as provided in logic programming systems such as B-Prolog \cite{Zhou08tab}, XSB \cite{Swift11}, YAP \cite{Costa11}, and Mercury \cite{Somogyi06}, has been shown to be a viable declarative language construct for describing dynamic programming solutions for various kinds of real-world applications, ranging from program analysis, parsing, deductive databases, theorem proving, model checking, to logic-based probabilistic learning. The main idea of tabling is to memorize the answers to subgoals in a table area and use the answers to resolve their variant or subsumed descendants. This idea of caching previously calculated solutions, called {\it memoization}, was first used to speed up the evaluation of functions \cite{Michie68}. Tabling can get rid of not only infinite loops for bounded-term-size programs but also redundant computations in the execution of recursive programs. While Datalog programs require tabling only subgoals with atomic arguments, many other programs such as those dealing with complex language corpora or bio-sequences require tabling structured data. Unfortunately, none of the current tabling systems can process structured data satisfactorily. Consider, for example, the predicate {\tt is\_list/2}:
\begin{verbatim}
    :-table is_list/1.
    is_list([]).
    is_list([_|L]):-is_list(L).
\end{verbatim}
For the subgoal {\tt is\_list([1,2,...,N])}, the current tabled Prolog systems demonstrate a higher complexity than linear in {\tt N}: B-Prolog (version 7.6 and older) consumes linear space but quadratic time; YAP, with a global trie for all tabled structured terms \cite{Raimundo11}, consumes linear space but quadratic time; XSB is quadratic in both time and space. The nonlinear complexity is due to the data structure used to represent tabled subgoals and answers and the need to copy terms into and from the table area.

The inefficiency of early versions of B-Prolog in handling large sequences has been reported and a program transformation method has been proposed to index ground structured data to work around the problem \cite{Have12}. In old versions of B-Prolog, tabled subgoals and answers were organized as hash tables, and {\it input sharing} was exploited to allow a tabled subgoal to share its ground structured arguments with its answers and its descendant subgoals. Input sharing enabled B-Prolog to consume only linear space for the tabled subgoal {\tt is\_list([1,2,...,N])}. Nevertheless, since the hash code was based on the first three elements of a list, the time complexity for a query like {\tt is\_list([1,1,...,1])} was quadratic in the length of the list. B-Prolog didn't support {\it output sharing}, i.e. letting different answers share structured data. Therefore, on the tabled version of the permutation program that generates all permutations through backtracking, B-Prolog would create $n\times n!$ cons cells where $n$ is the length of the given list.

This problem with tabling structured data has been noticed before and several remedies have been attempted. One well known technique used in parsing is to represent sentences as position indexed facts rather than lists. XSB provides tabled grammar predicates that convert list representation to position representation by redefining the built-in predicate {\tt 'C'/3}.\footnote{Personal communication with David S. Warren, 2011.} The position representation is also used for PCFG parsing in PRISM \cite{Sato08}. A program transformation method has been  proposed to index ground structured data to work around the quadratic time complexity of B-Prolog's tabling system \cite{Have12}. Nevertheless, these remedies have their limitations: the position representation disallows natural declarative modeling of sequences and the program transformation incurs considerable overhead. Have and Christiansen advocate for native support of data sharing in tabled Prolog systems for better scalability of their bio-sequence analysis application \cite{Have12}.

We have implemented full data sharing in B-Prolog in response to the manifesto. In the new version of B-Prolog, both input sharing and output sharing are exploited to allow tabled subgoals and answers to share ground structured data. {\it Hash-consing} \cite{ERSHOV58},  a technique originally used in functional programming to share values that are structurally equal \cite{Goto74,Appel93}, is adopted to memorize structured data in the table area. This technique avoids storing the same ground term more than once in the table area. While hash-consing can reduce the space consumption when sharing is effective, it does not change the time complexity. To avoid the extra linear time factor in dealing with sequences, we enhance hash-consing with input sharing and hash code memoization. For each compound term, an extra cell is used to store its hash code. 

Our main contribution in this paper is to apply hash-consing to tabling and enhance it with techniques to make it time efficient. The resulting system  demonstrates linear complexity in terms of both space and time on the query {\tt is\_list(L)} for any kind of ground list {\tt L}. As another contribution, we also compare tries with hash consing in the tabling context. As long as sequences are concerned, a trie allows for sharing of prefixes while hash-consing allows for sharing of ground suffixes. While we can build examples that arbitrarily favor one over the other, for recursively defined predicates such as {\tt is\_list}, it is more common for subgoals to share suffixes than prefixes. The enhanced hash-consing greatly improves the scalability of PRISM on sequence analysis applications. Our experimental results on a simulator of a hidden Markov model show that PRISM with enhanced hash-consing is asymptotically better than the previous version that supports no hash-consing.

The remainder of the paper is structured as follows: Section \ref{sec:primitives} defines the primitive operations on the table area used in a typical tabling system; Section \ref{sec:hash} presents the hash tables for subgoals and answers, and describes the copy algorithm for copying data from the stack/heap to the table area;  Section \ref{sec:hash-consing} modifies the copy algorithm to accommodate hash-consing; Section \ref{sec:enhanced} describes the techniques for speeding up computation of hash codes; Section \ref{sec:eval} evaluates the new tabling system with enhanced hash-consing; Section \ref{sec:related} gives a survey of related work; and Section \ref{sec:con} concludes the paper.

\section{\label{sec:primitives}Operations on the Table Area}
A tabling system uses a data area, called {\it table area}, to store tabled subgoals and their answers. A tabling system, whether it is suspension-based SLG \cite{david:slg} or iteration-based linear tabling \cite{Zhou08tab}, relies on the following three primitive operations to access and update the table area.\footnote{The interpretation of these operations may vary depending on implementations.}

\begin{description}
\item[Subgoal lookup and registration:] This operation is used when a tabled subgoal is encountered in execution. It looks up the subgoal table to see if there is a variant of the subgoal. If not, it inserts the subgoal (termed a {\it pioneer} or {\it generator}) into the subgoal table. It also allocates an answer table for the subgoal and its variants. Initially, the answer table is empty. If the lookup finds that there already is a variant of the subgoal in the table, then the record stored in the table is used for the subgoal (called a {\it consumer}). Generators and consumers are dealt with differently. In linear tabling, for example, a generator is resolved using clauses and a consumer is resolved using answers; a generator is iterated until the fixed point is reached and a consumer fails after it exhausts all the existing answers.

\item[Answer lookup and registration:] This operation is executed when a clause succeeds in generating an answer for a tabled subgoal. If a variant of the answer already exists in the table, it does nothing; otherwise, it inserts the answer into the answer table for the subgoal. When the lazy consumption strategy (also called local strategy) is used, a failure occurs no matter whether the answer is in the table or not, which drives the system to produce the next answer. 

\item[Answer return:] When a consumer is encountered, an answer is returned immediately if any. On backtracking, the next answer is returned. A generator starts consuming its answers after it has exhausted all its clauses. Under the lazy consumption strategy, a top-most looping generator does not return any answer until it is complete.
\end{description}

\section{\label{sec:hash}Hash Tables for Subgoals and Answers}
The data structures used for the table area are orthogonal to the tabling mechanism, whether it is suspension-based or iteration-based; they can be hash tables, tries, or some other data structures. In this section, we consider hash tables and the operations for the table area without data sharing. 

A hash table, called a {\it subgoal table},  is used for all tabled subgoals. For each tabled subgoal and its variants, there is a record in the subgoal table, which includes, amongst others, the following fields:

\begin{tabular}{ll} \hline
{\tt AnswerTable}: & Pointer to the answer table for the subgoal \\ 
{\tt sym}: & The functor of the subgoal \\ 
{\tt A1...An}: & The arguments the subgoal \\ \hline
\end{tabular}
When a tabled predicate is invoked by a subgoal, the subgoal table is looked up to see if a variant of the subgoal exists. If not, a record is allocated and the arguments are copied from the stack/heap to the table area. The copy of the subgoal shares no structured terms with the original subgoal and all of its variables are numbered so that they have different identities from those in the original subgoal.

The record of a subgoal in the subgoal table includes a pointer to another hash table, called an {\it answer table}, for storing answers produced for the subgoal. For each answer and its variants, there is a record in the answer table, which stores amongst others a pointer to a copy of the answer.  When an answer is produced for a subgoal, the subgoal's answer table is looked up to see if a variant of the answer exists. If not, a record is allocated and the answer is copied from the stack/heap to the table area. The answers in a subgoal's answer table are connected from the oldest one to the newest one such that they can be consumed by the subgoal one by one through backtracking.

In the implementation, a hash table is represented as an array. To add an item into a hash table, the system computes the hash code  of the item and uses the hash code modulo the size of the array to determine a slot for the item. All items hashed to the same slot are connected as a linked list, called a {\it hash chain}. A hash table is expanded when the number of records in it exceeds the size of the array. 

The WAM representation \cite{Warren83} is used to represent both terms on the heap and terms in the table area except that variables in tabled terms are numbered. A term is represented by a word containing a value and a tag. The tag distinguishes the type of the term. It may be {\tt REF} denoting a reference, {\tt ATM} an atomic value, {\tt STR} a structure, {\tt LST} a cons, or {\tt NUMVAR} a numbered variable. A {\tt STR}-tagged reference to a structure $f(t_1,\ldots,t_n)$ points to a block of $n+1$ consecutive words where the first word points to the functor $f/n$ in the symbol table and the remaining $n$ words store the $n$ components of the structure. An {\tt LST}-tagged reference to a list cons $[H|T]$ points to a block of two consecutive words where the first word stores the car $H$ and the second word stores the cdr $T$.   

Figure \ref{fig:copy} gives the definition of the function {\tt copy\_term} that copies a numbered term from the stack/heap to the table area. The hash function is designed in such a way that the hash code of a non-ground term is always 0. The function call {\tt seq\_hcode(code1,code2)} gives the combined hash code of the two hash codes from two components:
\begin{verbatim}
    int seq_hcode(int code1, int code2){
        if (code1==0) return 0;
        if (code2==0) return 0;
        return code1+31*code2+1;
    }
\end{verbatim}
If either code is 0, then the resulting code is 0 too.\footnote{Note that this way of combing hash codes is for hash consing terms. For the subgoal and answer tables, hash codes are combined in a different way.}

It is assumed that all the variables in a subgoal have been numbered before the arguments are copied. In the real implementation, variables are numbered inside the function {\tt copy\_term}. The function call {\tt copy\_subgoal\_args(src,des,arity)} copies the arguments of a numbered subgoal to the table area where {\tt (src-i)} points to the {\tt i}th argument on the stack and {\tt (des+i)} is the destination in the table area where the argument is copied to. In the TOAM architecture \cite{Zhou12} on which B-Prolog is based, arguments are passed through the stack and the stack grows downward from high addresses to low ones. That is why {\tt (src-1)} points to the first argument and {\tt (src-arity)} points to the last argument of the subgoal. A similar function is used to copy answers to the table area.

\begin{figure}[t]
\begin{center}
\begin{scriptsize}
\begin{tabbing}
aa \= aaa \= aaa \= aaa \= aaa \= aaa \= aaa \kill
{\tt int copy\_subgoal\_args(TermPtr src, TermPtr des, int arity)\{ } \\
\> {\tt hcsum = 0; } \\
\> {\tt for (i=1;i<=arity;i++)\{} \\
\> \> {\tt hcode = copy\_term(*(src-i), des+i); } \\
\> \> {\tt hc\_sum = seq\_hcode(hc\_sum,hcode);} \\
\> {\tt \}} \\
\> {\tt return hc\_sum; } \\
{\tt \}} \\
\\
{\tt int copy\_term(Term t, TermPtr des)\{ } \\
\> {\tt deref(t); } \\
\> {\tt switch (tag(t))\{ } \\
\> {\tt case NUMVAR: } \\
\> \> {\tt *des = t; } \\
\> \> {\tt return 0; } \\
\> {\tt case ATM: } \\
\> \> {\tt *des = t; } \\
\> \> {\tt return atomic\_hcode(t); } \\
\> {\tt case LST: } \\
\> \> {\tt p1 = untag(t);} \\
\> \> {\tt p2 = allocate\_from\_table(2); } \\
\> \> {\tt car\_code = copy\_term(*p1, p2); } \\
\> \> {\tt cdr\_code = copy\_term(*(p1+1), p2+1); } \\
\> \> {\tt hcode = seq\_hcode(car\_code,cdr\_code);} \\
\> \> {\tt t1 = add\_tag(p2,LST); } \\
\> \> {\tt *des = t1;} \\
\> \> {\tt return hcode;} \\
\> {\tt case STR: } \\
\> \> {\tt p1 = untag(t);} \\
\> \> {\tt sym = *p1; } \\
\> \> {\tt arity = get\_arity(sym);} \\
\> \> {\tt p2 = allocate\_from\_table(arity+1); } \\
\> \> {\tt hcode = *p2 = sym; } \\
\> \> {\tt for (i=1;i<=arity;i++)} \\
\> \> \> {\tt hcode = seq\_hcode(hcode, copy\_term(*(p1+i), p2+i)); } \\
 \> \>{\tt t1 = add\_tag(p2,STR); } \\
\> \> {\tt *des = t1; } \\
\> \> {\tt return hcode;} \\
\> {\tt \} /* end switch */}\\
{\tt \} /* end copy\_term */} 
\end{tabbing}
\end{scriptsize}
\end{center}
\caption{\label{fig:copy}Copy data to the table area with no sharing.}
\end{figure}

The function {\tt copy\_term} is not tail recursive and can easily cause the native C stack to overflow when copying large lists. In the real implementation, an iterative version is used to copy a list and compute its hash code. For a cons, the function needs to compute the hash codes of the car and the cdr before computing its hash code. The function does this in two passes: in the first pass it reverses the list and in the second pass it computes the hash codes while reversing the list back.

The function {\tt copy\_term} exploits no sharing of data. Consider, for example, the following program and the query {\tt is\_list([1,2])}. After completion of the query, the subgoal table contains three tabled subgoals, {\tt is\_list([1,2])}, {\tt is\_list([2])}, and {\tt is\_list([])}, and each subgoal's answer table contains an answer that is just a copy of the subgoal itself. No data are shared among the copies of the terms. So there are two separate copies of {\tt [1,2]} and two separate copies of {\tt [2]} in the table area. In the WAM representation of lists, a cons requires two words to store, so 12 words are used in total. In general, the query {\tt is\_list([1,2,...,N])} consumes $O({\tt N}^2)$ space in the table area.

\section{\label{sec:hash-consing}Hash-Consing of Ground Compound Terms}
Hash-consing, like tabling, is a memoization technique which uses a hash table to memorize values that have been created. Before creating a new value, it looks up the table to see if the value exists. If so, it reuses the existing value, otherwise, it inserts the value into the table. The concept of hash-consing originates from implementations of Lisp that attempt to reuse cons cells that have been constructed before \cite{Goto74}. This technique has also been suggested for Prolog (e.g., for sharing answers of findall/3 \cite{OKeefe01}), but its use in Prolog implementations is unknown, not to mention its use in tabling.

Let's call the hash table used for all ground terms {\it terms-table}. Figure \ref{fig:copy2} gives an updated version of {\tt copy\_term} that performs hash-consing. If the term is a list or a structure, the function copies it into the table area first. If the term is ground, it then calls the function {\tt hash\_consing(t1,hcode)} to look up the terms-table to see if a copy of {\tt t1} already exists in the table. If so, {\tt hash\_consing(t1,hcode)} returns the copy; otherwise, it inserts {\tt t1} into the terms-table and returns {\tt t1} itself. If an old copy in the terms-table is returned (\verb+t1 != t2+), the function deallocates the memory space allocated for the current copy.

With hash-consing, the query {\tt ?-is\_list([1,2])} only creates one copy of {\tt [1,2]} in the table area and the list is shared by the subgoals and the answers. As {\tt [2]} is the cdr of {\tt [1,2]}, no separate copy is stored for it. So, only 4 words are used in total for the list. The number of words used for hashing the two lists varies, depending on if there is a collision. If no collision occurs, two slots in the terms-table are used; otherwise, one slot in the terms-table is used and one node with two words is used to chain the two lists. So in the worst case, 7 words are needed in total. 

\begin{figure}[t]
\begin{center}
\begin{scriptsize}
\begin{tabbing}
aa \= aaa \= aaa \= aaa \= aaa \= aaa \= aaa \kill
{\tt int copy\_term(Term t, TermPtr des)\{ } \\
\> {\tt deref(t); } \\
\> {\tt switch (tag(t))\{ } \\
\> {\tt case NUMVAR: } \\
\> \> {\tt *des = t; } \\
\> \> {\tt return 0; } \\
\> {\tt case ATM: } \\
\> \> {\tt *des = t; } \\
\> \> {\tt return atomic\_hcode(t); } \\
\> {\tt case LST: } \\
\> \> {\tt p1 = untag(t);} \\
\> \> {\tt p2 = allocate\_from\_table(2); } \\
\> \> {\tt car\_code = copy\_term(*p1, p2); } \\
\> \> {\tt cdr\_code = copy\_term(*(p1+1), p2+1); } \\
\> \> {\tt hcode = seq\_hcode(car\_code,cdr\_code);} \\
\> \> {\tt t1 = add\_tag(p2,LST); } \\
\> \>  {\tt \color{red} if (is\_ground\_hcode(hcode))\{}\\
\> \> \> {\tt  \color{red} t2 = hash\_consing(t1,hcode); } \\
\> \> \> {\tt  \color{red} if (t1 != t2)\{} \\
\> \> \> \> {\tt  \color{red} deallocate\_to\_table(2); } \\
\> \> \> \> {\tt  \color{red} t1 = t2;} \\
\> \> \> {\tt  \color{red} \}} \\
\> \> {\tt  \color{red} \}} \\
\> \> {\tt *des = t1;} \\
\> \> {\tt return hcode;} \\
\> {\tt case STR: } \\
\> \> {\tt p1 = untag(t);} \\
\> \> {\tt sym = *p1; } \\
\> \> {\tt arity = get\_arity(sym);} \\
\> \> {\tt p2 = allocate\_from\_table(arity+1); } \\
\> \> {\tt hcode = *p2 = sym; } \\
\> \> {\tt for (i=1;i<=arity;i++)} \\
\> \> \> {\tt hcode = seq\_hcode(hcode, copy\_term(*(p1+i), p2+i)); } \\
 \> \>{\tt t1 = add\_tag(p2,STR); } \\
 \> \>{\tt \color{red} if (is\_ground\_hcode(hcode))\{} \\
 \> \> \>{\tt \color{red}  t2 = hash\_consing(t1,hcode); } \\
 \> \> \>{\tt \color{red}   if (t1 != t2)\{} \\
 \> \> \> \>{\tt \color{red}   deallocate\_to\_table(arity+1); } \\
 \> \> \> \>{\tt \color{red}   t1 = t2;} \\
 \> \> \>{\tt \color{red}   \}} \\
 \> \>{\tt \color{red}  \} } \\
\> \> {\tt *des = t1; } \\
\> \> {\tt return hcode;} \\
\> {\tt \} /* end switch */}\\
{\tt \} /* end copy\_term */} 
\end{tabbing}
\end{scriptsize}
\end{center}
\caption{\label{fig:copy2}Copy data with hash-consing.}
\end{figure}

\section{\label{sec:enhanced}Enhanced Hash-Consing}
With hash-consing, the tabled subgoal {\tt is\_list([1,...,N])} consumes only linear table space now. Nevertheless, its time complexity remains quadratic in {\tt N}. This is because for each descendant subgoal {\tt is\_list([K,...,N])} ({\tt K}$>$1) the hash code of the list {\tt [K,...,N]} has to be computed and the terms-table has to be looked up. We enhance hash-consing with two techniques to lower the time complexity of {\tt is\_list([1,...,N])} to linear.\footnote{The worst case time complexity is still quadratic in theory if a poorly designed hash function is used.}

\subsection{Hash code memoization}
The first technique is to table hash codes of structured terms in the table area. For each structure or a list cons in the table area, we use an extra word to store its hash code. The WAM representation of terms is not changed. The word for the hash code of a compound term is located right before the term.  So assume {\tt p} is the untagged reference to a structure or a list cons, then {\tt p-1} references the hash code.

Figure \ref{fig:copy3} gives a new version of {\tt copy\_term} that tables hash codes. Tabled hash codes are used for two purposes. Firstly, when searching for the term {\tt t1} in the hash chain, the function {\tt hash\_consing(t1,hcode)} always compares the hash codes first and only when the codes are equal will it compare the terms. Secondly, the  system reuses the tabled hash codes of terms when it expands a hash table and rehashes the terms into the new hash table.

With tabled hash codes, the subgoal {\tt is\_list([1,...,N])} still takes quadratic time since the list {\tt [1,...,N]} resides on the heap and for each descendant subgoal, the hash code of the argument is not available and hence has to be computed. To avoid this computation, we introduce input sharing.

\begin{figure}[t]
\begin{center}
\begin{scriptsize}
\begin{tabbing}
aa \= aaa \= aaa \= aaa \= aaa \= aaa \= aaa \kill
{\tt int copy\_term(Term t, TermPtr des)\{ } \\
\> {\tt deref(t); } \\
\> {\tt switch (tag(t))\{ } \\
\> {\tt case NUMVAR: } \\
\> \> {\tt *des = t; } \\
\> \> {\tt return 0; } \\
\> {\tt case ATM: } \\
\> \> {\tt *des = t; } \\
\> \> {\tt return atomic\_hcode(t); } \\
\> {\tt case LST: } \\
\> \> {\tt p1 = untag(t);} \\
\> \> {\tt \color{red}  if (!is\_heap\_reference(p1))\{ }\\
\> \> \> {\tt \color{red} *des = t;} \\
\> \> \> {\tt \color{red} return *(p1-1); /* return the tabled hash code */} \\
\> \> {\tt \color{red}\}} \\
\> \> {\tt p2 = allocate\_from\_table({\color{red} 3});} \\
\> \> {\tt \color{red} p2++;} \\
\> \> {\tt car\_code = copy\_term(*p1, p2); } \\
\> \> {\tt cdr\_code = copy\_term(*(p1+1), p2+1); } \\
\> \> {\tt hcode = seq\_hcode(car\_code,cdr\_code);} \\
\> \> {\tt \color{red} *(p2-1) = hcode;} \\
\> \> {\tt t1 = add\_tag(p2,LST); } \\
\> \>  {\tt if (is\_ground\_hcode(hcode))\{}\\
\> \> \> {\tt  t2 = hash\_consing(t1,hcode); } \\
\> \> \> {\tt  if (t1 != t2)\{} \\
\> \> \> \> {\tt  deallocate\_to\_table({\color{red} 3}); } \\
\> \> \> \> {\tt  t1 = t2;} \\
\> \> \> {\tt  \}} \\
\> \> {\tt  \}} \\
\> \> {\tt *des = t1;} \\
\> \> {\tt return hcode;} \\
\> {\tt case STR: } \\
\> \> {\tt p1 = untag(t);} \\
\> \> {\tt \color{red}  if (!is\_heap\_reference(p1))\{ }\\
\> \> \> {\tt \color{red} *des = t;} \\
\> \> \> {\tt \color{red} return *(p1-1); /* return the tabled hash code */} \\
\> \> {\tt \color{red}\}} \\
\> \> {\tt sym = *p1; } \\
\> \> {\tt arity = get\_arity(sym);} \\
\> \> {\tt p2 = allocate\_from\_table({\color{red} arity+2}); } \\
\> \> {\tt \color{red} p2++;} \\
\> \> {\tt hcode = *p2 = sym; } \\
\> \> {\tt for (i=1;i<=arity;i++)} \\
\> \> \> {\tt hcode = seq\_hcode(hcode, copy\_term(*(p1+i), p2+i)); } \\
\> \> {\tt \color{red} *(p2-1) = hcode;} \\
 \> \>{\tt t1 = add\_tag(p2,STR); } \\
 \> \>{\tt if (is\_ground\_hcode(hcode))\{} \\
 \> \> \>{\tt  t2 = hash\_consing(t1,hcode); } \\
 \> \> \>{\tt   if (t1 != t2)\{} \\
 \> \> \> \>{\tt   deallocate\_to\_table({\color{red} arity+2}); } \\
 \> \> \> \>{\tt   t1 = t2;} \\
 \> \> \>{\tt   \}} \\
 \> \>{\tt  \} } \\
\> \> {\tt *des = t1; } \\
\> \> {\tt return hcode;} \\
\> {\tt \} /* end switch */}\\
{\tt \} /* end copy\_term */} 
\end{tabbing}
\end{scriptsize}
\end{center}
\caption{\label{fig:copy3}Tabling hash codes while copying with hash-consing.}
\end{figure}

\subsection{Input Sharing}
Input sharing amounts to letting a subgoal share its ground terms with its answers and descendant subgoals. Consider the tabled subgoal {\tt is\_list([1,2,3])}. The answer is the same as the subgoal, so it shares the term {\tt [1,2,3]} with the subgoal in the table area. The direct descendant subgoal is {\tt is\_list([2,3])}. Since the list {\tt [2,3]} is a suffix of {\tt [1,2,3]}, the descendant subgoal should share it with the original subgoal in the table area.

To implement input sharing, we let the copying procedure set the frame slot of an argument of a tabled subgoal to the address of the copied argument in the table area if the argument is a ground structured term. So for the tabled subgoal {\tt is\_list([1,2,3])}, the frame slot of the argument initially references the list {\tt [1,2,3]} on the heap. After the subgoal is copied to the table area, the frame slot is set to reference the copy of the list in the table area. In this way, the list will be shared by answers and the descendant subgoals. For programs that do not use destructive assignments, which is the case for tabled programs, updating frame slots this way causes no problem.

The function {\tt copy\_subgoal\_args} shown in Figure \ref{fig:input} implements input sharing. When an argument is found to be ground, the function lets the stack slot of the argument reference its copy in the table area. The function {\tt copy\_term} (in Figure \ref{fig:copy3}) tests the reference to a compound term to see if the term needs to be copied. If it is not a heap reference, then the referenced term must reside in the table area and thus can be reused.

Note that our input sharing scheme has its limitation in the sense that it fails to facilitate sharing of ground components in non-ground arguments. Consider, for example, the subgoal {\tt is\_list([X,2,3])}. The suffix {\tt [2,3]} will not be  shared through input sharing in our implementation since the argument is not ground. It will eventually be shared through hash-consing, but its hash code needs to be computed again when it occurs in a descendant subgoal or an answer.

\begin{figure}[t]
\begin{center}
\begin{scriptsize}
\begin{tabbing}
aa \= aaa \= aaa \= aaa \= aaa \= aaa \= aaa \kill
{\tt int copy\_subgoal\_args(TermPtr src, TermPtr des, int arity)\{ } \\
\> {\tt hcsum = 0; } \\
\> {\tt for (i=1;i<=arity;i++)\{} \\
\> \> {\tt hcode = copy\_term(*(src-i), des+i); } \\
\> \> {\tt \color{red} if (is\_ground\_hcode(hcode)) *(src-i) = *(des+1);} \\
\> \> {\tt hc\_sum = seq\_hcode(hc\_sum,hcode);} \\
\> {\tt \}} \\
\> {\tt return hc\_sum; } \\
{\tt \}} \\
\end{tabbing}
\end{scriptsize}
\end{center}
\caption{\label{fig:input}Input sharing by updating frame slots.}
\end{figure}

\section{\label{sec:eval}Evaluation}
The improved tabling system described in this paper has been implemented and made available with B-Prolog version 7.7 (BP7.7). We evaluate the proposed approach by comparing BP7.7 with YAP (version 6.3.2) and XSB (version 3.3.6), and also the previous version of B-Prolog, version 7.6 (BP7.6), which did not have enhanced hash-consing. We also compare it with indexed programs produced by the transformation proposed in \cite{Have12} running on B-Prolog 7.6 ({\it indexed}). We use the {\tt is\_list/1} predicate, the {\tt edit\_distance/3}\footnote{The source code is available in \cite{Have12}.} program, and a PRISM program to show the effectiveness of the proposed techniques. We also test on a program that favors prefix sharing with tries more than suffix sharing with hash-consing. In addition, we also show results for the CHAT suite and the ATR parser, the traditional benchmarks used to evaluate tabling systems.

The results are obtained on a Linux machine with 16 2.4 GHz, 64 bit Intel Xeon(R) E7340 processor cores and 64 GB of memory. For this evaluation, only a single processor core is utilized. CPU times (in seconds) and table space (in kilobytes) consumptions are measured using the \texttt{statistics/1} built-in for BP and XSB, and {\tt table\_statistics/1} for YAP.

Table \ref{islist_repeat} shows the results on the query \texttt{is\_list([1,1,...,1])} where N is the number of 1s in the list. All the systems except for BP7.6 demonstrate a close-to-linear complexity. The higher time complexity of BP7.6 is due to that fact that BP7.6 only uses the first three elements of a list as the key and hashing degenerates into linear search for the query because of hash collision. The difference in time among BP7.7, YAP and XSB is at least a large constant factor. As mentioned above, a trie allows for sharing of prefixes while hash-consing allows for sharing of suffixes as long as lists are concerned. For a list that contains repeated data, there are an equal number of prefixes and suffixes, and hence both types of sharing are equally favored. The difference between BP7.7 and {\it indexed} is only a small constant factor.

\begin{table}
\caption{\label{islist_repeat}Results on {\tt is\_list([1,1,...,1])}}
\begin{tabular}{ |l|r|r|r|r|r|r|r|r|r|r| }
\cline{1-11}
 & \multicolumn{2}{c|}{BP7.7} & \multicolumn{2}{c|}{BP7.6} & \multicolumn{2}{c|}{\it indexed} & \multicolumn{2}{c|}{YAP} & \multicolumn{2}{c|}{XSB} \\ \cline{1-11} 
N & \multicolumn{1}{c|}{time} & \multicolumn{1}{c|}{space} & \multicolumn{1}{c|}{time} & \multicolumn{1}{c|}{space} & \multicolumn{1}{c|}{time} & \multicolumn{1}{c|}{space} & \multicolumn{1}{c|}{time} & \multicolumn{1}{c|}{space} & \multicolumn{1}{c|}{time} & \multicolumn{1}{c|}{space} \\ \cline{1-11} 
 500 & 0.000 &  33 &  0.098 &  43 & 0.001 &  39 & 0.007 &  90 & 0.003 &  399 \\ \cline{1-11} 
1000 & 0.001 &  66 &  0.776 &  86 & 0.003 &  78 & 0.033 & 180 & 0.010 &  567 \\ \cline{1-11} 
1500 & 0.001 &  99 &  2.608 & 128 & 0.004 & 117 & 0.073 & 269 & 0.019 &  735 \\ \cline{1-11} 
2000 & 0.002 & 131 &  6.169 & 171 & 0.005 & 156 & 0.134 & 359 & 0.037 &  903 \\ \cline{1-11} 
2500 & 0.001 & 164 & 12.034 & 214 & 0.006 & 195 & 0.186 & 449 & 0.058 & 1071 \\ \cline{1-11} 
3000 & 0.002 & 197 & 20.777 & 257 & 0.008 & 234 & 0.282 & 539 & 0.078 & 1239 \\ \cline{1-11} 
3500 & 0.002 & 229 & 32.975 & 300 & 0.009 & 273 & 0.384 & 629 & 0.108 & 1407 \\ \cline{1-11} 
4000 & 0.003 & 264 & 49.204 & 343 & 0.011 & 312 & 0.498 & 719 & 0.139 & 1575 \\ \cline{1-11} 
4500 & 0.003 & 297 & 70.048 & 386 & 0.011 & 351 & 0.571 & 809 & 0.177 & 1743 \\ \cline{1-11} 
5000 & 0.003 & 330 & 96.112 & 429 & 0.013 & 390 & 0.729 & 898 & 0.217 & 1911 \\ \cline{1-11} 
\end{tabular}
\end{table}

Table \ref{islist_random} shows the results on the query {\tt is\_list(L)} where {\tt L} is a list of random constants.\footnote{A random number generator is used to generate the lists. For each size, the same list was used for all the systems.} BP consumes linear space and linear time; YAP consumes linear space thanks to the global trie for terms but takes quadratic time; XSB is quadratic in both time and space. For random lists, suffix sharing with hash consing is clearly more effective than prefix sharing with tries.

\begin{table}
\caption{\label{islist_random}Results on \texttt{is\_list(L)} where {\tt L} contains random data.} 
\begin{tabular}{ |l|r|r|r|r|r|r|r|r|r|r| }
\cline{1-11}
 & \multicolumn{2}{c|}{BP7.7} & \multicolumn{2}{c|}{BP7.6} & \multicolumn{2}{c|}{\it indexed} & \multicolumn{2}{c|}{YAP} & \multicolumn{2}{c|}{XSB} \\ \cline{1-11} 
N & \multicolumn{1}{c|}{time} & \multicolumn{1}{c|}{space} & \multicolumn{1}{c|}{time} & \multicolumn{1}{c|}{space} & \multicolumn{1}{c|}{time} & \multicolumn{1}{c|}{space} & \multicolumn{1}{c|}{time} & \multicolumn{1}{c|}{space} & \multicolumn{1}{c|}{time} & \multicolumn{1}{c|}{space} \\ \cline{1-11} 
 500 & 0.000 &  33 &  0.000 &  43 & 0.002 &  39 & 0.008 &  90 & 0.024 &   9990 \\ \cline{1-11} 
1000 & 0.001 &  66 &  0.001 &  86 & 0.002 &  78 & 0.032 & 180 & 0.063 &  39236 \\ \cline{1-11} 
1500 & 0.001 &  99 &  0.001 & 128 & 0.004 & 117 & 0.082 & 270 & 0.142 &  87991 \\ \cline{1-11} 
2000 & 0.001 & 132 &  0.002 & 171 & 0.005 & 156 & 0.134 & 360 & 0.252 & 156269 \\ \cline{1-11} 
2500 & 0.001 & 164 &  0.003 & 214 & 0.007 & 195 & 0.218 & 450 & 0.387 & 244071 \\ \cline{1-11} 
3000 & 0.002 & 197 &  0.003 & 257 & 0.008 & 234 & 0.341 & 540 & 0.559 & 351401 \\ \cline{1-11} 
3500 & 0.002 & 229 &  0.004 & 300 & 0.010 & 273 & 0.401 & 630 & 0.766 & 478260 \\ \cline{1-11} 
4000 & 0.003 & 264 &  0.005 & 343 & 0.011 & 312 & 0.537 & 719 & 0.978 & 624640 \\ \cline{1-11} 
4500 & 0.003 & 297 &  0.006 & 386 & 0.012 & 351 & 0.703 & 809 & 1.244 & 790555 \\ \cline{1-11} 
5000 & 0.004 & 330 &  0.008 & 429 & 0.013 & 390 & 0.894 & 899 & 1.504 & 975990 \\ \cline{1-11} 
\end{tabular}
\end{table}

Tables \ref{edit_repeat}  and \ref{edit_random} show the results on the {\tt edit\_distance} program with repeated data and random data, respectively. The main predicate {\tt edit(L1,L2,D)} in the program computes the distance between {\tt L1} and {\tt L2}, i.e., the number of substitutions, insertions and deletions needed to transform {\tt L1} to {\tt L2}. The tabled version finds all solutions. BP7.7 is significantly faster than BP7.6 on the type of queries that use repeated data. BP7.7 also outperforms YAP and XSB in both time and space on both types of queries. Similar to the {\tt is\_list} benchmark, enhanced hash-consing is asymptotically more effective than tries on random data.

\begin{table}
\caption{\label{edit_repeat}Results on \texttt{edit([1,1,...,1],[1,1,...,1],D).}}
\begin{tabular}{ |l|r|r|r|r|r|r|r|r|r|r| }
\cline{1-11}
 & \multicolumn{2}{c|}{BP7.7} & \multicolumn{2}{c|}{BP7.6} & \multicolumn{2}{c|}{\it indexed} & \multicolumn{2}{c|}{YAP} & \multicolumn{2}{c|}{XSB} \\ \cline{1-11} 
N & \multicolumn{1}{c|}{time} & \multicolumn{1}{c|}{space} & \multicolumn{1}{c|}{time} & \multicolumn{1}{c|}{space} & \multicolumn{1}{c|}{time} & \multicolumn{1}{c|}{space} & \multicolumn{1}{c|}{time} & \multicolumn{1}{c|}{space} & \multicolumn{1}{c|}{time} & \multicolumn{1}{c|}{space} \\ \cline{1-11} 
 30 & 0.000 &   60 &    0.026 &   97 & 0.003 &   90 & 0.005 &   213 &  0.006 &   1273 \\ \cline{1-11} 
 60 & 0.003 &  233 &    0.726 &  378 & 0.016 &  348 & 0.034 &   819 &  0.057 &   4341 \\ \cline{1-11} 
 90 & 0.007 &  519 &    5.189 &  841 & 0.036 &  776 & 0.107 &  1820 &  0.235 &   9435 \\ \cline{1-11} 
120 & 0.015 &  917 &   21.216 & 1487 & 0.064 & 1372 & 0.266 &  3214 &  0.736 &  16554 \\ \cline{1-11} 
150 & 0.022 & 1427 &   63.536 & 2316 & 0.102 & 2137 & 0.517 &  5002 &  1.635 &  25698 \\ \cline{1-11} 
180 & 0.031 & 2051 &  156.072 & 3328 & 0.142 & 3071 & 0.942 &  7183 &  3.041 &  36868 \\ \cline{1-11} 
210 & 0.047 & 2786 &  334.190 & 4523 & 0.208 & 4173 & 1.533 &  9759 &  5.035 &  50064 \\ \cline{1-11} 
240 & 0.060 & 3634 &  646.550 & 5900 & 0.267 & 5445 & 2.367 & 12728 &  7.662 &  65285 \\ \cline{1-11} 
270 & 0.074 & 4595 & 1159.182 & 7460 & 0.339 & 6885 & 3.081 & 16090 & 11.327 &  82531 \\ \cline{1-11} 
300 & 0.095 & 5668 & 1955.331 & 9204 & 0.448 & 8493 & 4.401 & 19847 & 15.664 & 101803 \\ \cline{1-11} 
\end{tabular}
\end{table}

\begin{table}
\caption{\label{edit_random}Results on \texttt{edit(L1,L2,D)} where {\tt L1} and {\tt L2} contain random data.}
\begin{tabular}{ |l|r|r|r|r|r|r|r|r|r|r| }
\cline{1-11}
 & \multicolumn{2}{c|}{BP7.7} & \multicolumn{2}{c|}{BP7.6} & \multicolumn{2}{c|}{\it indexed} & \multicolumn{2}{c|}{YAP} & \multicolumn{2}{c|}{XSB} \\ \cline{1-11} 
N & \multicolumn{1}{c|}{time} & \multicolumn{1}{c|}{space} & \multicolumn{1}{c|}{time} & \multicolumn{1}{c|}{space} & \multicolumn{1}{c|}{time} & \multicolumn{1}{c|}{space} & \multicolumn{1}{c|}{time} & \multicolumn{1}{c|}{space} & \multicolumn{1}{c|}{time} & \multicolumn{1}{c|}{space} \\ \cline{1-11} 
 30 & 0.001 &   61 & 0.000 &   97 & 0.004 &   90 & 0.005 &   214 &  0.011 &    4148 \\ \cline{1-11} 
 60 & 0.003 &  234 & 0.006 &  378 & 0.020 &  348 & 0.045 &   822 &  0.099 &   27706 \\ \cline{1-11} 
 90 & 0.010 &  521 & 0.016 &  841 & 0.038 &  776 & 0.118 &  1823 &  0.313 &   89645 \\ \cline{1-11} 
120 & 0.017 &  919 & 0.033 & 1487 & 0.067 & 1372 & 0.298 &  3218 &  0.759 &  209183 \\ \cline{1-11} 
150 & 0.027 & 1430 & 0.057 & 2316 & 0.105 & 2137 & 0.591 &  5007 &  1.501 &  404752 \\ \cline{1-11} 
180 & 0.038 & 2054 & 0.094 & 3328 & 0.148 & 3071 & 1.058 &  7190 &  2.771 &  695363 \\ \cline{1-11} 
210 & 0.056 & 2790 & 0.156 & 4523 & 0.217 & 4173 & 1.695 &  9766 &  4.271 & 1099906 \\ \cline{1-11} 
240 & 0.073 & 3639 & 0.219 & 5900 & 0.282 & 5445 & 2.687 & 12736 &  6.247 & 1637354 \\ \cline{1-11} 
270 & 0.092 & 4600 & 0.297 & 7460 & 0.352 & 6885 & 3.782 & 16100 &  8.787 & 2327276 \\ \cline{1-11} 
300 & 0.114 & 5674 & 0.435 & 9204 & 0.466 & 8493 & 5.248 & 19857 & 11.954 & 3187340 \\ \cline{1-11} 
\end{tabular}
\end{table}

Table \ref{prism_hmm} compares BP7.7 and BP7.6 on the PRISM program that simulates a two-state hidden Markov model \cite{prism:website}. For our benchmarking purpose, the training data of the form {\tt hmm([a,b,a,b,...])} are used, and only the time and space required to find all the explanations are measured. While BP7.7 consumes slightly more space than BP7.6 due to the overhead of hash-consing, it outperforms BP7.6 in time by a linear factor.

\begin{table}
\caption{\label{prism_hmm} Results on the PRISM program HMM.}
\begin{tabular}{ |l|r|r|r|r|}

\cline{1-5}
 & \multicolumn{2}{c|}{BP7.7} & \multicolumn{2}{c|}{BP7.6} \\ \cline{1-5} 
N & \multicolumn{1}{c|}{time} & \multicolumn{1}{c|}{space} & \multicolumn{1}{c|}{time} & \multicolumn{1}{c|}{space} \\ \cline{1-5} 
 2000 & 0.002 &  222 &   1.164 & 179 \\ \cline{1-5} 
 3000 & 0.005 &  333 &   3.911 & 269 \\ \cline{1-5} 
 4000 & 0.006 &  444 &   9.249 & 359 \\ \cline{1-5} 
 5000 & 0.008 &  555 &  18.044 & 449 \\ \cline{1-5} 
 6000 & 0.010 &  666 &  31.150 & 539 \\ \cline{1-5} 
 7000 & 0.011 &  776 &  49.441 & 628 \\ \cline{1-5} 
 8000 & 0.013 &  889 &  73.774 & 718 \\ \cline{1-5} 
 9000 & 0.015 & 1000 & 105.049 & 808 \\ \cline{1-5} 
10000 & 0.018 & 1111 & 144.140 & 898 \\ \cline{1-5} 
\end{tabular}
\end{table}

Although it is more common for subgoals of recursive programs to share suffixes than prefixes, it is possible to find programs on which prefix sharing with tries is more effective than suffix sharing with hash-consing. The following gives such a program:
\begin{verbatim}
    :-table create_list/2.
    create_list(N,L):-
        between(1,N,I),
        range(1,I,L).
\end{verbatim}
The query {\tt create\_list(N,L)} creates {\tt N} lists {\tt [1]}, {\tt [1,2]}, ..., and {\tt [1,2,...,N]} that have only common prefixes. As shown in Table \ref{tab:clist}, XSB consumes linear space, while BP and YAP consume quadratic space. YAP tables all suffixes into the global trie for terms and there are $O({\tt N}^2)$ suffixes. BP7.7 consumes more table space than BP7.6 since all the terms are hash-consed but none is shared. BP7.6 is slower than BP7.7 since the hash function used in BP7.6, which is based on the first three elements of a list, results in more collisions than BP7.7.

\begin{table}
\caption{\label{tab:clist}Results on {\tt create\_list(N,L)}.}
\begin{tabular}{ |l|r|r|r|r|r|r|r|r|r|r| }
\cline{1-9}
 & \multicolumn{2}{c|}{BP7.7} & \multicolumn{2}{c|}{BP7.6} & \multicolumn{2}{c|}{YAP} & \multicolumn{2}{c|}{XSB} \\ \cline{1-9} 
N & \multicolumn{1}{c|}{time} & \multicolumn{1}{c|}{space} & \multicolumn{1}{c|}{time} & \multicolumn{1}{c|}{space} & \multicolumn{1}{c|}{time} & \multicolumn{1}{c|}{space} & \multicolumn{1}{c|}{time} & \multicolumn{1}{c|}{space} \\ \cline{1-9} 
 500 & 0.035 &   2417 &   0.107 &   990 & 0.039 &   3965 & 0.023 & 290 \\ \cline{1-9} 
1000 & 0.201 &   9564 &   0.827 &  3937 & 0.201 &  15742 & 0.043 & 348 \\ \cline{1-9} 
1500 & 0.654 &  21635 &   2.989 &  8831 & 0.523 &  35332 & 0.095 & 407 \\ \cline{1-9} 
2000 & 0.969 &  37926 &   7.245 & 15679 & 0.962 &  62734 & 0.169 & 465 \\ \cline{1-9} 
2500 & 2.151 &  60082 &  14.130 & 24480 & 1.699 &  97949 & 0.264 & 524 \\ \cline{1-9} 
3000 & 2.660 &  85890 &  24.343 & 35249 & 2.630 & 140976 & 0.378 & 583 \\ \cline{1-9} 
3500 & 3.276 & 116011 &  38.397 & 47956 & 3.739 & 191816 & 0.517 & 641 \\ \cline{1-9} 
4000 & 4.011 & 150192 &  57.217 & 62616 & 5.071 & 250468 & 0.675 & 700 \\ \cline{1-9} 
4500 & 7.319 & 194310 &  80.994 & 79229 & 6.978 & 316933 & 0.853 & 758 \\ \cline{1-9} 
5000 & 8.316 & 238885 & 110.631 & 97796 & 9.267 & 391211 & 1.051 & 817 \\ \cline{1-9} 
\end{tabular}
\end{table}

Table \ref{tab:classic} compares the systems on the CHAT benchmark suite and the ATR parser. There is almost no difference between BP7.7 and BP7.6 in time and the space overhead incurred by hash-consing is noticeable. Hash-consing has no positive effect on these programs because the sequences used in the programs are very short.

\begin{table}
\caption{\label{tab:classic}Results on the CHAT benchmarks and the ATR parser.}
\begin{tabular}{ |l|r|r|r|r|r|r|r|r|r| }
\cline{1-9}
 & \multicolumn{2}{c|}{BP 7.7} & \multicolumn{2}{c|}{BP 7.6} & \multicolumn{2}{c|}{YAP} & \multicolumn{2}{c|}{XSB} \\ \cline{1-9} 
Benchmark & \multicolumn{1}{c|}{time} & \multicolumn{1}{c|}{space} & \multicolumn{1}{c|}{time} & \multicolumn{1}{c|}{space} & \multicolumn{1}{c|}{time} & \multicolumn{1}{c|}{space} & \multicolumn{1}{c|}{time} & \multicolumn{1}{c|}{space} \\ \cline{1-9} 
cs\_o & 0.015 & 198 & 0.0129 & 11 & 0.009 & 26 & 0.011 & 285 \\ \cline{1-9} 
cs\_r & 0.025 & 332 & 0.026 & 11 & 0.019 & 27 & 0.022 & 286 \\ \cline{1-9} 
disj & 0.008 & 108 & 0.009 & 11 & 0.005 & 23 & 0.007 & 277 \\ \cline{1-9} 
gabriel & 0.011 & 111 & 0.012 & 9 & 0.006 & 20 & 0.008 & 272 \\ \cline{1-9} 
kalah & 0.008 & 90 & 0.008 & 15 & 0.006 & 35 & 0.008 & 304 \\ \cline{1-9} 
pg & 0.006 & 69 & 0.006 & 7 & 0.004 & 15 & 0.006 & 263 \\ \cline{1-9} 
read & 0.057 & 987 & 0.058 & 23 & 0.099 & 46 & 0.030 & 327 \\ \cline{1-9} 
atr & 0.509 & 15111 & 0.543 & 5947 & 0.325 & 52520 & 0.280 & 45400 \\ \cline{1-9} 
\end{tabular}
\end{table}

\section{\label{sec:related}Related Work}
Since structure sharing \cite{Boye72a} was discarded and the Warren Abstract Machine (WAM) \cite{Warren83} triumphed as the implementation model of Prolog, there has been little attention paid to exploiting data sharing in Prolog implementations.\footnote{A lot of work has been done on indexing Prolog terms, but indexing is a different kind of sharing since it does not consider reuse of terms from different sources.}  In his Diploma thesis \cite{Neumerkel89}, Ulrich Neumerkel gave several example Prolog programs that would consume an-order-of-magnitude less space with data sharing than without sharing. He proposed applying hash-consing and DFA-minimization to sharing terms including cyclic ones. The proposed approach would incur considerable overhead if every compound term is hash-consed when created, and hence it is infeasible to incorporate the approach into the WAM. Following Appel and Goncalves's hash-consing garbage collector for SML/NJ \cite{Appel93}, Nguyen and Demoen recently built a similar garbage collector for hProlog \cite{Nguyen}. The garbage collector hash-conses compound terms on the heap in one phase and performs absorption in another phase such that for the replications of a compound term only one copy is kept and all the others are garbage collected. Their experiment basically confirms the disappointing result reported in Appel and Goncalves's paper: the overhead outweighs the gain except for special programs. 

Hash-consing can be applied to the built-in predicate {\tt findall/3}, as suggested by O'Keefe  \cite{OKeefe01}, to avoid repeatedly copying the same term in different answers. Currently, B-Prolog is the only Prolog system that supports hash-consing for {\tt findall/3}. It employs a hash table for ground terms in the findall area. The algorithm and memory manager developed for the table area is reused for the findall area. With hash-consing, the system copies a ground term only once when copying answers from the findall area to the heap. Input sharing is exploited in the same way as for tabled subgoals. For a {\tt findall} call, the compiler converts it into a call to a temporary predicate such that each argument of the generator occupies one slot in the stack frame. At runtime, the system first copies the arguments of the generator from the stack/heap to the findall area before the generator is executed. When an argument of the generator is found to be a ground compound term, its frame slot is set to reference the copy in the findall area. In this way, the argument and its subterms can be reused by the answers and the descendant calls. Nguyen and Demoen's implementation of input sharing for {\tt findall/3} \cite{Nguyen} distinguishes between old terms that are created before the generator and new terms that are generated by the generator, and have answers share the old terms. Their scheme can exploit sharing of not only ground arguments but also ground terms in non-ground arguments. Their scheme may not be suited for tabled data since, unlike data in the findall area which live and die with the generator, tabled data are permanent. Also, their implementation does not exploit output sharing.

A trie has been a popular data structure for organizing tabled subgoals and answers \cite{Ram98}. It is adopted by all the tabled Prolog systems except B-Prolog. As far as lists are concerned, a trie facilitates sharing of the prefixes while hash-consing allows for sharing of the suffixes. So for the two lists {\tt [1,2]} and {\tt [1,2,3]}, the former shares the same path as the latter in the trie, but they are treated as separate lists when hash-consed; for the two lists {\tt [2,3]} and {\tt [1,2,3]}, however, a trie allows for no sharing while hash-consing allows for complete sharing.

Another advantage of tries is that they can be used to perform both variant testing and subsumption testing, and thus can be used in both variant-based and subsumption-based tabling systems. Hash-consing, on the other hand, can be used to perform equivalence testing only and thus cannot directly be used for subsumption-based tabling.

Terms stored in a trie have a different representation from terms on the heap. For example, in the YAP system, tries are represented as trie instructions \cite{Costa11}. For this reason, when an answer is returned, it must be copied from its trie in the table area to the heap even if it is ground. In our system, structured ground terms in the table area have exactly the same representation as on the heap, so when they occur in an answer they do not need to be copied when the answer is returned. 

In the original implementation of XSB and YAP, one trie is used for all tabled subgoals, and for each subgoal one trie is used for the answer table.  To enhance sharing, Raimundo and Rocha propose using a global trie for all tabled structured terms \cite{Raimundo11}. Due to the necessity of copying answers from the table area to the heap, the time complexity remains the same even when the space complexity drops.

To some extent, the idea of representing sentences as position indexed facts \cite{Have12,xsb} is similar to hash-consing in the sense that a hash-consed term always is associated with a hash code. The translation from a program that deals with sequences represented as lists into one that uses position representation is not trivial. When difference lists are involved, the translation is even more complicated. The program obtained after translation may lose sharing opportunities. Therefore, hash-consing is a more practical solution to sharing than program transformation.

As far as we know, our implementation is the first attempt to apply hash-consing to tabling. Our implementation enhances hash-consing with input sharing and hash code memoization to speed-up computation of hash codes. The extra cell used to store the hash code of a compound term is overhead if the term is never shared. Nevertheless, while the increase of space is always a constant factor, the gain in speed can be linear in the size of the data. 

\section{\label{sec:con} Conclusion}
We have presented an implementation of hash-consing for tabling structured data. Hash-consing facilitates sharing of structured data and can eliminate the extra linear factor of space complexity commonly seen in early tabling systems when dealing with sequences. Hash-consing alone does not change the time complexity. We have enhanced it with input sharing and hash code memoization to eliminate the extra linear factor of time complexity in dealing with sequences. The resulting tabling system significantly improves the scalability of language parsing and bio-sequence analysis applications. 

Our work will shed some light on the discussion on what data structure to use for tabled data. A trie is suitable for sharing prefixes and hash-consing is suitable for sharing suffixes of sequences. Although it is possible to find programs that make prefix sharing arbitrarily better than suffix sharing, it is more common for subgoals of recursive programs to share suffixes than prefixes. Therefore, hash-consing is in general a better choice than tries as a data structure for representing tabled data. Hash-consing as it is in our implementation is not suitable for subsumption-based tabling. It is future work to adapt hash-consing to subsumption testing.

\section*{Acknowledgements}
The PRISM system has been the motivation for this project and we thank Taisuke Sato and Yoshitaka Kameya for their discussion.  We also thank the anonymous referees for their detailed comments on the presentation. Neng-Fa Zhou was supported in part by NSF (No.1018006) and Christian Theil Have was supported by the project ¡°Logic-statistic modelling and analysis of biological sequence data¡± funded by the NABIIT program under the Danish Strategic Research Council.


\begin{thebibliography}{}

\bibitem[\protect\citeauthoryear{Appel and de~Rezende~Goncalves}{Appel and
  de~Rezende~Goncalves}{2003}]{Appel93}
{\sc Appel, A.~W.} {\sc and} {\sc de~Rezende~Goncalves, M.~J.} 2003.
\newblock Hash-consing garbage collection.
\newblock Technical Report TR 74-03, Princeton University.

\bibitem[\protect\citeauthoryear{Boyer and Moore}{Boyer and
  Moore}{1972}]{Boye72a}
{\sc Boyer, R.~S.} {\sc and} {\sc Moore, J.~S.} 1972.
\newblock A sharing of structure in theorem proving programs.
\newblock {\em Machine Intelligence\/}~{\em 7}, 101--116.

\bibitem[\protect\citeauthoryear{Chen and Warren}{Chen and
  Warren}{1996}]{david:slg}
{\sc Chen, W.} {\sc and} {\sc Warren, D.~S.} 1996.
\newblock Tabled evaluation with delaying for general logic programs.
\newblock {\em Journal of the ACM\/}~{\em 43,\/}~1, 20--74.

\bibitem[\protect\citeauthoryear{Ershov}{Ershov}{1959}]{ERSHOV58}
{\sc Ershov, A.} 1959.
\newblock On programming of arithmetic operations.
\newblock {\em Communications of the ACM\/}~{\em 1,\/}~8, 3--6.

\bibitem[\protect\citeauthoryear{Goto}{Goto}{1974}]{Goto74}
{\sc Goto, E.} 1974.
\newblock Monocopy and associative algorithms in extended {Lisp}.
\newblock Technical Report TR 74-03, University of Tokyo.

\bibitem[\protect\citeauthoryear{Have and Christiansen}{Have and
  Christiansen}{2012}]{Have12}
{\sc Have, C.~T.} {\sc and} {\sc Christiansen, H.} 2012.
\newblock Efficient tabling of structured data using indexing and program
  transformation.
\newblock In {\em {PADL}}. LNCS 7149, 93--107.

\bibitem[\protect\citeauthoryear{Michie}{Michie}{1968}]{Michie68}
{\sc Michie, D.} 1968.
\newblock ``memo'' functions and machine learning.
\newblock {\em Nature\/}, 19--22.

\bibitem[\protect\citeauthoryear{Neumerkel}{Neumerkel}{1989}]{Neumerkel89}
{\sc Neumerkel, U.} 1989.
\newblock Garbage collection in {Prolog} systems (in {German}).
\newblock Ph.D. thesis, Thesis, Technical University of Vienna.

\bibitem[\protect\citeauthoryear{Nguyen and Demoen}{Nguyen and
  Demoen}{2012}]{Nguyen}
{\sc Nguyen, P.-L.} {\sc and} {\sc Demoen, B.} 2012.
\newblock Representation sharing for {Prolog}.
\newblock {\em TPLP\/}.

\bibitem[\protect\citeauthoryear{O'Keefe}{O'Keefe}{2001}]{OKeefe01}
{\sc O'Keefe, R.~A.} 2001.
\newblock O(1) reversible tree navigation without cycle.
\newblock {\em TPLP\/}~{\em 1,\/}~5, 617--630.

\bibitem[\protect\citeauthoryear{Raimundo and Rocha}{Raimundo and
  Rocha}{2011}]{Raimundo11}
{\sc Raimundo, J.} {\sc and} {\sc Rocha, R.} 2011.
\newblock Global trie for subterms.
\newblock In {\em CICLOPS}.

\bibitem[\protect\citeauthoryear{Ramakrishnan, Rao, Sagonas, Swift, and
  Warren}{Ramakrishnan et~al\mbox{.}}{1998}]{Ram98}
{\sc Ramakrishnan, I.}, {\sc Rao, P.}, {\sc Sagonas, K.}, {\sc Swift, T.}, {\sc
  and} {\sc Warren, D.} 1998.
\newblock Efficient access mechanisms for tabled logic programs.
\newblock {\em Journal of Logic Programming\/}~{\em 38}, 31--54.

\bibitem[\protect\citeauthoryear{{Santos Costa}, Rocha, and Damas}{{Santos
  Costa} et~al\mbox{.}}{2012}]{Costa11}
{\sc {Santos Costa}, V.}, {\sc Rocha, R.}, {\sc and} {\sc Damas, L.} 2012.
\newblock The {YAP Prolog} system.
\newblock {\em TPLP, Special Issue on {Prolog} Systems\/}~{\em 12,\/}~1-2,
  5--34.

\bibitem[\protect\citeauthoryear{Sato and Kameya}{Sato and
  Kameya}{2008}]{Sato08}
{\sc Sato, T.} {\sc and} {\sc Kameya, Y.} 2008.
\newblock New advances in logic-based probabilistic modeling by {PRISM}.
\newblock In {\em Probabilistic Inductive Logic Programming}. 118--155.

\bibitem[\protect\citeauthoryear{Sato, Zhou, Kameya, and Yizumi}{Sato
  et~al\mbox{.}}{2010}]{prism:website}
{\sc Sato, T.}, {\sc Zhou, N.-F.}, {\sc Kameya, Y.}, {\sc and} {\sc Yizumi, Y.}
  2010.
\newblock The {PRISM} user's manual.
\newblock {http://www.mi.cs.titech.ac.jp/prism/}.

\bibitem[\protect\citeauthoryear{Somogyi and Sagonas}{Somogyi and
  Sagonas}{2006}]{Somogyi06}
{\sc Somogyi, Z.} {\sc and} {\sc Sagonas, K.} 2006.
\newblock Tabling in {Mercury}: Design and implementation.
\newblock In {\em {PADL}}. LNCS 3819, 150--167.

\bibitem[\protect\citeauthoryear{Swift and Warren}{Swift and
  Warren}{2012}]{Swift11}
{\sc Swift, T.} {\sc and} {\sc Warren, D.~S.} 2012.
\newblock {XSB}: Extending {Prolog} with tabled logic programming.
\newblock {\em TPLP, Special issue on Prolog systems\/}~{\em 12,\/}~1-2,
  157--187.

\bibitem[\protect\citeauthoryear{Swift, Warren, et~al\mbox{.}}{Swift
  et~al\mbox{.}}{2009}]{xsb}
{\sc Swift, T.}, {\sc Warren, D.~S.}, {\sc et~al\mbox{.}} 2009.
\newblock {The XSB Programmer's Manual: vols. 1 and 2}.
\newblock http://xsb.sf.net.

\bibitem[\protect\citeauthoryear{Warren}{Warren}{1983}]{Warren83}
{\sc Warren, D. H.~D.} 1983.
\newblock An abstract {P}rolog instruction set.
\newblock Technical note 309, SRI International.

\bibitem[\protect\citeauthoryear{Zhou}{Zhou}{2012}]{Zhou12}
{\sc Zhou, N.-F.} 2012.
\newblock The language features and architecture of {B-Prolog}.
\newblock {\em TPLP, Special Issue on {Prolog} Systems\/}~{\em 12,\/}~1-2,
  189--218.

\bibitem[\protect\citeauthoryear{Zhou, Sato, and Shen}{Zhou
  et~al\mbox{.}}{2008}]{Zhou08tab}
{\sc Zhou, N.-F.}, {\sc Sato, T.}, {\sc and} {\sc Shen, Y.-D.} 2008.
\newblock Linear tabling strategies and optimizations.
\newblock {\em TPLP\/}~{\em 8,\/}~1, 81--109.

\end{thebibliography}
\end{document}